\documentclass[showpacs,twocolumn,prl]{revtex4}
\usepackage{amsmath,graphicx}
\usepackage{color,ulem}

\definecolor{darkred}{rgb}{0.90,0,0}
\definecolor{darkgreen}{rgb}{0,0.60,.2}
\definecolor{darkblue}{rgb}{0,0,1}
\definecolor{grey}{cmyk}{0,0,0,0.25}
\definecolor{orange}{cmyk}{0,0.6,0.8,0}

\begin{document}

\title{Microwave 
response of an NS ring coupled to a superconducting resonator}
\author{F. Chiodi,$^{1}$ M. Ferrier,$^{1}$ K.Tikhonov,$^{2}$ P. Virtanen,$^{3}$ T.T. Heikkil\"a,$^{3}$ M. Feigelman,$^{2}$ S. Gu\'eron$^{1}$ and H.Bouchiat$^{1}$ }
\affiliation{$^{1}$ LPS, Univ. Paris-Sud, CNRS, UMR 8502, F-91405 Orsay Cedex, France,$^2$ L. D. Landau Institute for Theoretical Physics, Kosygin str.2, Moscow 119334, Russia,$^3$Low Temperature Laboratory, Aalto University, P.O. Box 15100, FI-00076 Aalto, Finland. }

\begin{abstract}
 A  long phase coherent normal (N) wire between superconductors (S) is characterized by   a dense phase dependent Andreev spectrum . We probe this spectrum in a  high frequency phase biased  configuration,  by coupling an  NS  ring  to a  multimode superconducting  resonator. We  detect  a dc flux and frequency  dependent response whose dissipative  and non dissipative  components are related by a simple Debye relaxation law with a characteristic time of the order of the diffusion time through the N part of the ring.  The flux dependence  exhibits   $h/2e$  periodic oscillations with a large harmonics content at temperatures where the Josephson current is  purely sinusoidal.  This is  explained considering that the populations of the Andreev levels are frozen on the time-scale of the experiments.
\end{abstract}
\maketitle
Most properties of a non superconducting (normal) metal connected to two superconductors (an SNS junction) can be seen as resulting from the Andreev states (AS) in the normal metal, and the occupation of those states. Andreev states are correlated electron-hole  eigenstates in the normal metal, determined by boundary conditions imposed by the superconducting banks. In a long diffusive normal metal (of length $L$ grater than the superconducting coherence length $\xi_s$), the AS spectrum is a quasi continuum of levels with a small energy gap  $E_g $ \cite{heikkila,spivak}. This so-called minigap depends solely on $L$ and the diffusion constant $D$, via the Thouless  energy $E_{th}=\hbar/\tau_D $ where $\tau_D = L^2/D$ is the diffusion time along the junction. $E_g$ is fully modulated by the phase difference $\varphi$ between the superconducting order parameters on both sides. $E_g (\varphi)$ is maximal at $\varphi=0$ with 
 $E_g (0)\simeq 3.1 E_{Th}$ and goes linearly to zero at $\varphi= \pi$, as was recently measured by scanning tunneling spectroscopy \cite{lesueur}. The phase dependent Josephson current $ I_J(\varphi)$ at equilibrium  sums the contributions  of each Andreev state of energy $\epsilon_n$, via $i_n=\frac{2e}{\hbar}\frac{ \partial\epsilon_n}{\partial \varphi}$, the current carried by  the  level $n$  of  occupation factor $p_n$.
 \begin{equation}
   I_J(\varphi) =\sum_n p_n(\epsilon_n (\varphi)) i_n (\varphi) 
 \end{equation}
 
  As was measured recently by Hall magnetometry \cite{strunk09}, the phase dependence of the supercurrent is non sinusoidal at low temperature, but turns sinusoidal at $T > E_g$, with an amplitude that decreases roughly exponentially with temperature on the scale of $E_g$. 

Whereas the equilibrium properties of the proximity effect (PE) in long SNS junction are well understood theoretically and experimentally, its dynamics is a  more complex and yet unresolved issue.  In particular  the  current response of the system  to a  time dependent phase  which affects both energy levels and their populations still needs  to be determined.  We expect at least the existence of  two time constants, one related to the relaxation of the populations to a change of the energy levels, another one related to the dynamics of the  Andreev levels, and possibly also a contribution due to the generation of quasiparticles. Which of the electron-electron, electron-phonon,   diffusion  or dephasing times are relevant in these processes? Experimentally, there are many ways in which to impose an out of equilibrium situation. Because of the Josephson relation linking the voltage to the time derivative of the phase, a voltage bias across the junction causes a time dependent phase, and the $I(V$) curve is a probe of the dynamics of the PE. Irradiating the junction with an rf field, which produces Shapiro steps in the $I(V)$ curve, is another. Both the critical current  in current biased experiments \cite{Warlaumont79, chiodi09} and the complete Josephson current phase relation \cite{strunk09} have been shown to be very sensitive to RF excitation at frequencies of the order of or larger than $1/\tau_D$. In addition, the electron-phonon scattering rate has also been shown to set the frequency scale for the hysteresis in current biased experiments \cite{chiodi09}.

In contrast to all these experiments performed in strongly non linear regimes, in this Letter we present a linear response experiment in which we measure both the non dissipative (in–phase), $\chi'$ and dissipative (out of phase), $\chi''$ current response of an SN ring. We measure these with imposing an oscillating phase difference, at several RF frequencies, as a function of the dc superconducting phase difference. Our results identify a current relaxation mechanism with a relaxation time of the order of the diffusion time. The temperature is greater than the minigap, so that the dc current phase relation is sinusoidal. Nevertheless, we find that the dissipative and the non dissipative components contain several harmonics. This shows on the one hand that at high frequency, the non dissipative response is not simply the flux derivative of the supercurrent, and also that dissipation is enhanced when the minigap closes.

 The experiment consists in inductively coupling an NS  ring to a  multimode superconducting  resonator operating between 300 MHZ and 6 GHZ.  In the linear response regime the  dc flux   dependences of $\chi'$ and  $\chi''$  are deduced  from the flux induced variations of the resonator's  resonances frequency and quality factor Q. A similar  technique based on single mode resonators  was already used for the investigation of short Josephson  junctions inbedded in  superconducting rings \cite{raskin,Illichev}.

 \begin{figure}
    \includegraphics[clip=true,width=7cm]{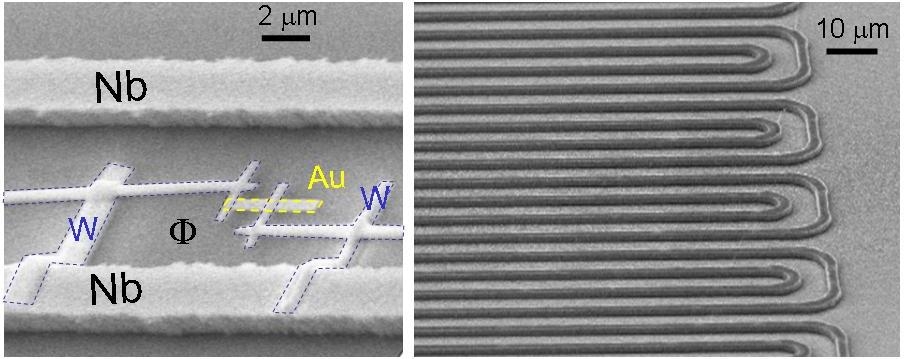}

\caption{Electron micrograph of the NS ring inserted in the  niobium resonator. }
    \label{Fig1}
\end{figure}

The  resonator consists of two parallel superconducting Nb meander lines ($2 \mu$m wide, $L_R=20$ cm long, and $4 \mu$m
apart, see Fig. \ref{Fig1}) patterned on a sapphire substrate [12]. One of these lines is weakly coupled to a RF generator via a small  on chip capacitance whose value is adjusted in order to preserve the  high Q of the resonator(80 000 at 20 mK), the other one is grounded.  
The resonance conditions are $L_R = \frac{n\lambda}{2}$, where the mode index n is an integer, and $\lambda$ the electromagnetic wavelength. The fundamental resonance frequency ($n=1$) is of the order of 360 MHz. The resonator is enclosed in a gold plated stainless-steel box, shielding it from electromagnetic noise, and cooled down to mK temperature. The high Q factors allows us to detect extremely small variations: $\frac{\delta f}{f} \sim \delta Q^{-1} \sim 10^{-9}$. It therefore provides  very accurate ac impedance measurements of mesoscopic objects.  This  technique was previously employed in  contactless measurements of the response  of Aharonov Bohm rings  \cite{deblock2002}. A single NS ring gives more signal than $10^4$ normal rings, for two reasons: First, the supercurrent in an NS ring is $g$ times larger than the persistent current in a normal ring of the same size as the N wire (g is the  conductance of the normal wire in $2e^2/h$ units). Second, the length of the S part  can be adjusted to optimize the inductive coupling between the ring and the resonator.    

The N part of the NS ring is a mesoscopic Au wire ($99.9999 \%$ purity, 1.2 x 0.38 $\mu m^2$, 50 $nm$ thick, normal state resistance ($R_N= 1/G_N =0.52$ $ \Omega$) corresponding to $g=2\times 10^4$), positionned between the two meander lines by e-beam lithography, and thermally evaporated onto the saphire substrate. We  estimate the  Thouless energy for this sample  to be $E_{Th} = 90\pm 10$ $mK$. The $S$  part of the ring is constructed by conecting the Au wire  to one of  the resonator lines   via  W nanowires deposited using a focused  Ga ion beam (FIB) to decompose  a tungstene carbonyle vapor. The superconducting transition temperature  of the W wires is 4K, their critical current is  1 mA,  and critical field is greater than 7 T \cite {kasumov06}.   This technique provides a good interface quality, thanks to a slight etching step prior to   deposition of W. We have checked that SNS junctions fabricated using this technique exhibit supercurrents and Shapiro steps comparable to long SNS junctions made with  more conventional techniques  \cite{chioditobepub}. The sample is thus an ac-squid embedded in the resonator.  The dc superconducting phase difference $\varphi$ at the boundaries of the N wire is imposed by a  magnetic flux $\Phi_{dc}$ created by a magnetic field  applied perpendicularly to the ring plane: $\varphi =-2\pi\Phi_{dc}/\Phi_0$ where $\Phi_0=h/2e$ is the superconducting flux quantum. In addition,  an ac flux $\delta \Phi_{\omega}cos(\omega t)$ is generated by the ac current in the resonator. At low enough frequency the current in the SNS ring should follow adiabatically the oscillating flux, and the ac response should be entirely in phase, given by the flux derivative of the Josephson current $\partial I_J(\Phi_{dc})/ \partial \Phi $ , also called the inverse kinetic inductance. But at high frequency the  ac current $\delta I_\omega$ in the ring should have both an in-phase and  an out-of-phase response to the ac flux : $\delta I_\omega (\Phi_{dc}) = \chi'(\omega , \Phi_{dc}) \delta \Phi_\omega cos(\omega t) + \chi''(\omega, \Phi_{dc})\delta \Phi_\omega sin(\omega t)$. Here $\chi =\chi'+i\chi''$ is the complex susceptibility of the ring related to its complex impedance $Z$ by $\chi=i\omega/Z$. 
\begin{figure}
    \includegraphics[clip=true,width=9cm]{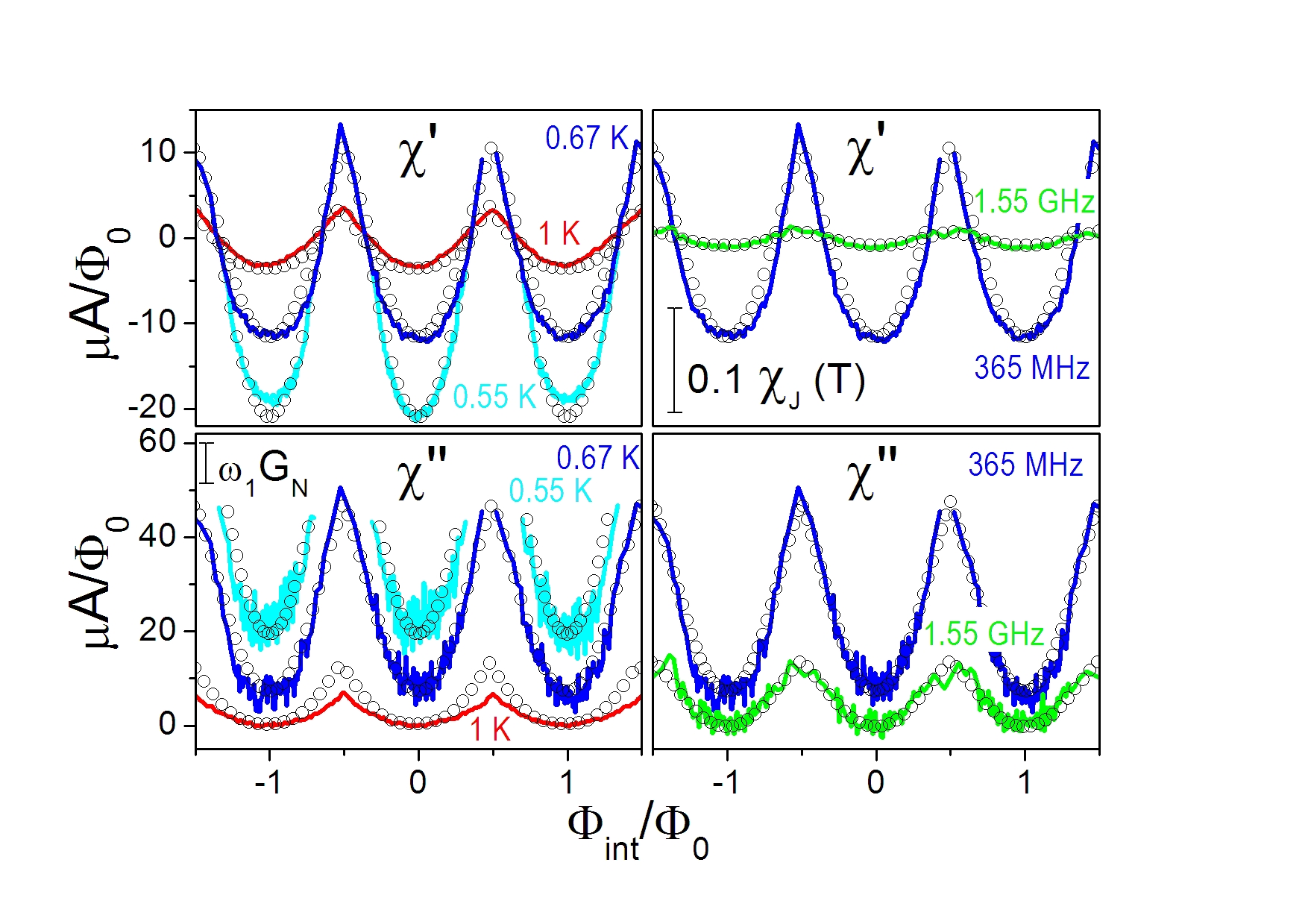}
    \caption{Flux dependence of the real $\chi'$ and imaginary susceptibility $\chi''$  for different temperatures    and  resonance frequencies  (left panel, f=365MHz, right panel T=0.67K).  The amplitude of the Josephson currents  calculated are $I_J(0.67~K) = 18~\mu A$ and $I_J(1K) = 6~\mu A$ . The data  are corrected from geometrical inductance effects, which induce  corrections  which are sizable at 0.67 and 0.55 K but negligible at 1K . Circles: fit with  $\chi_d (T)/ (1+ i\omega \tau)$ see expressions \ref{eqchi_in} and \ref{eqdebye2} calculated from Usadel equation \cite{virtanen} at  temperatures corresponding to $k_BT=5$, 6  and $9E_{Th}$ giving the best agreement with experimental data. The amplitude of the theoretical curves    has been rescaled by a factor 0.7. Note that this experiment only measures the flux dependence of $\chi$, all the curves have been arbitrarily shifted  along the vertical axis.} 
\label{Fig2}
\end{figure} 
The aim of the experiment is to determine $\chi (\Phi_{dc})$ of the  ring  at the  successive resonances of the resonator by measuring the  flux dependences of  the resonance frequencies and quality factor  which are related to $\chi(f_n)$ via:
\begin{equation}
\begin{array}{l}
\displaystyle\frac{\delta f_n(\Phi_{dc})}{f_n} = -\frac{1}{2}k_n\frac{M^2}{\cal{L}}\frac{\left[\chi'(1-L_g\chi') - L_g \chi''^2\right]}{D}\\
\displaystyle\delta\frac{ 1}{Q_n (\Phi_{dc})} = k_n\frac{M^2}{\cal{L}}\frac{\chi''}{D}\mbox{ with }
\displaystyle D=(1-L_g\chi')^2+L_g\chi''^2
\end{array}{}
 \label{perturb}
 \end{equation}
 $\cal{L}$ = 0.15 $\mu H$ is the  total inductance of the  resonator, $M\sim 5$ pH the  mutual inductance between the ring and the two lines of the resonator 
 and $L_g\sim 15$ pH the  ring's geometrical inductance.  The  coefficient $k_n$ accounts for the   spatial dependence of the ac field amplitude at frequency $f_n$.  
The limit $L_g=0$ yields the well known formulas where $\delta f/f$ and $\delta (1/Q)$ are respectively proportional to $ \chi'$ and $\chi''$.
 But since the  geometrical inductance of the ring  is finite, the internal dc  flux $\Phi_{int}$  differs from the applied flux $\Phi_{ext}$ and reads $\Phi_{int}=\Phi_{ext}+L_gI_J(\Phi_{int}) $. This screening effect leads to Eq.\ref{perturb} and to an hysteretic behavior at low temperature when the parameter $\beta=2\pi L_g I_c(T)/\Phi_0$ is larger than 1. (The critical current $I_c$ is the maximum amplitude of $I_J(\Phi)$). Because of this, it is not possible to access the whole current/phase relation  below 600 mK \cite{icdeT}.  
 
  In the following we present the dependences of $\chi$  upon the internal flux through the ring. Our main findings are summarized in Fig.2 showing the  flux oscillations of $\chi'$ and $\chi''$ for different temperatures and resonator modes. Each curve corresponds to an average of $30$ to $100$ magnetic field  scans.
Their  periodicity corresponds, as expected, to $\Phi_0$ through the ring.
  The first surprise is the observation, beside the non dissipative response $\chi'(\Phi)$, of a $larger$ dissipative response $\chi''(\Phi)$, over the entire frequency range investigated (365 MHz to 3 GHz). In addition, $\chi'(\Phi)$ and $\chi''(\Phi)$ are not
 harmonic functions of flux even though the dc Josephson current is sinusoidal in this temperature range. It is clear in particular that $\chi'(\Phi)$ is not equal to $\chi_J(\Phi)= \partial I_J(\Phi)/\partial\Phi$. 
  It is also interesting to compare the flux variation amplitudes $\delta \chi'$,  $\delta \chi''$ to the inverse kinetic inductance $\chi_J (T)= 2\pi I_C(T)/\Phi_0$ and to $\chi''_N = \omega G_N$ . Whereas we find that $\delta\chi'$ is  smaller than  $\chi_J(T)$, $\delta\chi''$ is  much larger than  $\chi''_N$.
 We now turn to the frequency dependence $\chi(\omega)$ in order to extract the relevant relaxation times. The amplitudes $\delta \chi'$ and $\delta \chi''$  are plotted in Fig.3 for the successive resonator eigen frequencies $f_n= \omega_n/2\pi$, at two temperatures. We find that the frequency dependence follows a simple Debye relaxation law, $\delta \chi(\omega) = \delta \chi (0)/(1+i\omega\tau)$, 
 with two adjustable parameters: $\delta \chi(0)$, of the order of $ \chi_J$, and the relaxation time $\tau$  equal to $0.6 \pm 0.2$ ns , which is 7 times the diffusion time. We could not detect significant temperature dependence of this time between 0.5 and 1K.

  So far we have discussed exclusively the linear response regime (microwave excitation powers between $10^{-15}$ and  $10^{-12}$ W). Figure 4 presents non linear effects in a larger NS ring with the same N wire. They show up at greater powers  correponding to an amplitude of the ac flux of the order of $\Phi_0$ through the ring: $\delta (1/Q)(\Phi)$ features sharp peaks at odd multiples of $\Phi_0/2$,  for which the minigap vanishes. The peaks widen with increasing microwave power. We attribute this extra dissipation to microwave induced Andreev pair breaking, which should occur preferentially in the range of dc flux where the minigap is the smallest. Raising temperature or frequency decreases the excitation amplitude threshold for the appearance of these peaks. The in phase response is also modified, with a striking increase of $\delta f(\Phi)$ in the whole flux range and a sign change of the flux dependence in the vicinity of $\Phi_0/2$. Non linearities also show up at lower excitation powers  when temperature is increased. This behavior recalls the strong modification of the dc current/phase relation under microwave irradiation  \cite{strunk09,heikkila10}.

 We now discuss  possible explanations of the linear-response data.
 One cause of dissipation in SNS junctions is the  relaxation of the population of  Andreev levels which  explains non equilibrium effects   in   voltage  biased configurations such as fractional Shapiro steps  \cite{argaman}. This  relaxation   is characterized by the inelastic time $\tau_{in}$  which is the shortest between the electron-electron  and the electron-phonon  scattering times. 
 It yields a frequency dependent contribution  to the response function $\chi$   proportional to the sum over the whole spectrum of the  square of the  single level current \cite{meso}.
\begin{equation}
\begin{array}{l}
\displaystyle \chi _{in}(\omega)= \Sigma _n \left[ p_n\frac{ \partial   i_n}{\partial\varphi} +   i_n(\varphi)\frac{\partial p_n}{\partial\epsilon_n }\frac{ \partial \epsilon_n} {\partial\varphi}\frac{1}{1+i\omega\tau_{in}}\right]\\ \displaystyle =\frac{\partial I_J(\varphi)}{\partial\varphi}-\frac{i\omega\tau_{in}}{1+i\omega\tau_{in}} F(\varphi,T).
\end{array}{}
\label{eqchi_in}
\end{equation}
where $F(\varphi,T)= -\Sigma_n \left[ i_n^2 \frac{\partial p_n}{\partial\epsilon_n } \right]$ can be written in the continuous spectrum limit in terms of the spectral current $J(\epsilon)$ and the density of states $\rho(\epsilon)$  of the Andreev levels as:
$ F(\varphi,T)= \int J^2(\varphi,\epsilon)/\left[k_BT\rho(\epsilon))\right]d\epsilon$.
The high frequency limit of$\chi_{in}$ is:
\begin{equation}
  \chi_d(\varphi,T) =\lim_{\omega \rightarrow \infty}  \chi _{in}(\omega) =\frac{\partial I_J(\varphi)}{\partial\varphi}-F(\varphi,T)
  \label{eqchid}
  \end{equation}
which can be also written  $\chi_d(\varphi,T)= \Sigma_n \left[ p_n \frac{\partial i_n}{\partial\varphi } \right]$ corresponding to time independent "frozen" populations of Andreev states.  We have calculated $F(\varphi,T)$ \cite{Limpitsky, virtanen} using Usadel equations \cite{usadel}. In the limit where $k_B T \gg E_{Th}$ it is  approximatively given by  $F(\varphi,T)= (2\pi G_N E_{Th} /e k_B T)(-\pi + \mod(\pi + \varphi, 2 \pi) - {\rm sgn}(\sin (\varphi))\sin ^2(\varphi/2)/\pi) \sin (\varphi) $.  
It is is nearly $\pi$ periodic with a sharp anomaly at $\varphi= \pi$ related to the closing of the minigap.  This leads to an important  harmonics content of $\chi(\varphi)$ at finite frequency just as found experimentally (see Fig.2) and discussed below. 
However the experimental flux dependence of $\delta \chi ''$  is similar to $\delta \chi'(\varphi)$ and  can  not be reproduced within this model.  
 \begin{figure}
    \includegraphics[clip=true,width=8cm]{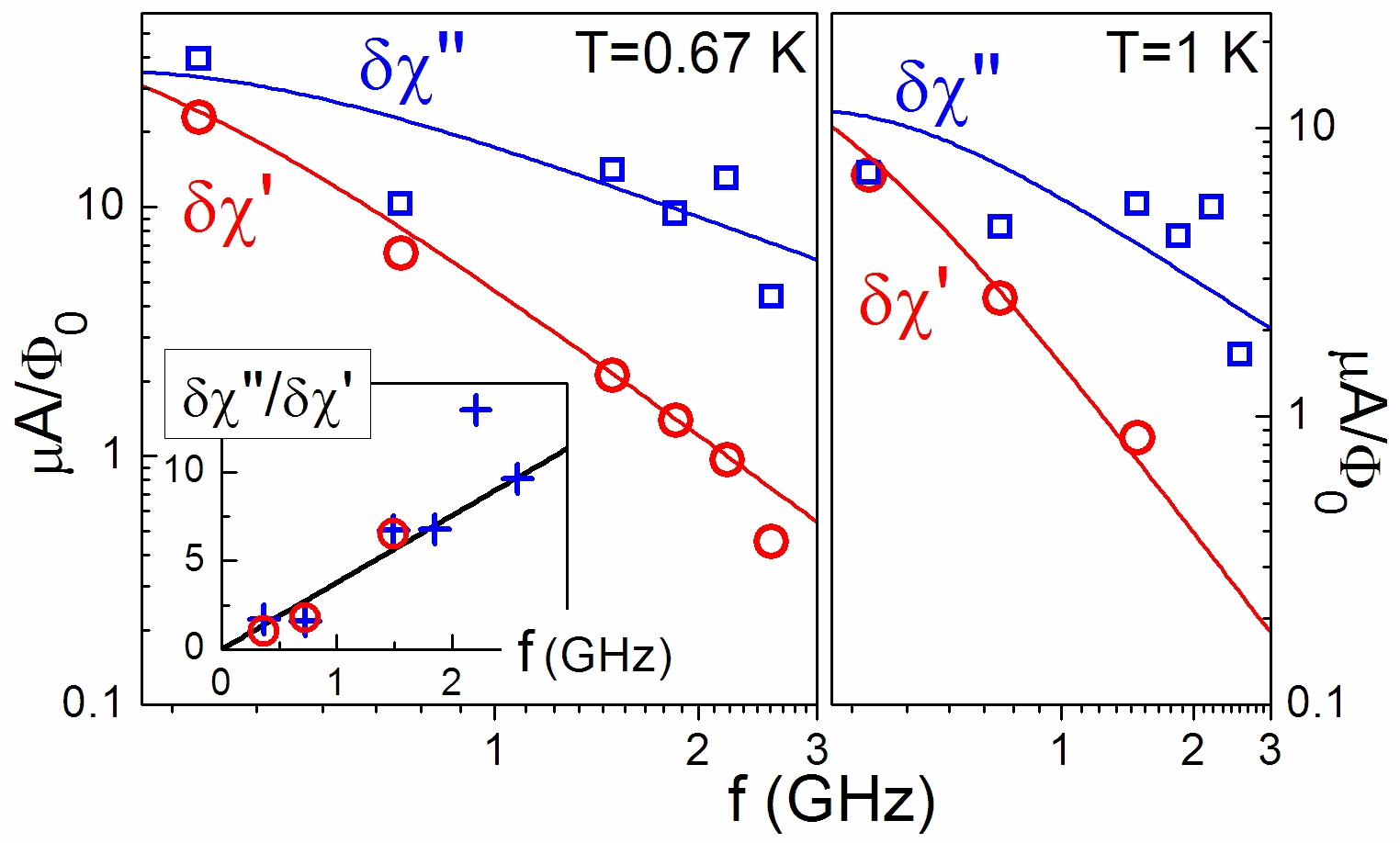}
    \caption{Frequency dependence of $\delta \chi'$ and $\delta \chi''$ for two different temperatures. Continuous lines: fits according  to Debye relaxation laws with $\tau =0.6$ $ ns$. Inset: frequency dependence of the ratio $\delta \chi''/\delta \chi'$ for several temperatures. The linear dependence in $2\pi f $ confirms the validity of the Debye relaxation fits. Circles: T=1$ $K, crosses: T=0.67$ $K }
    \label{Fig3}
\end{figure}  
\begin{figure}
    \includegraphics[clip=true,width=8cm]{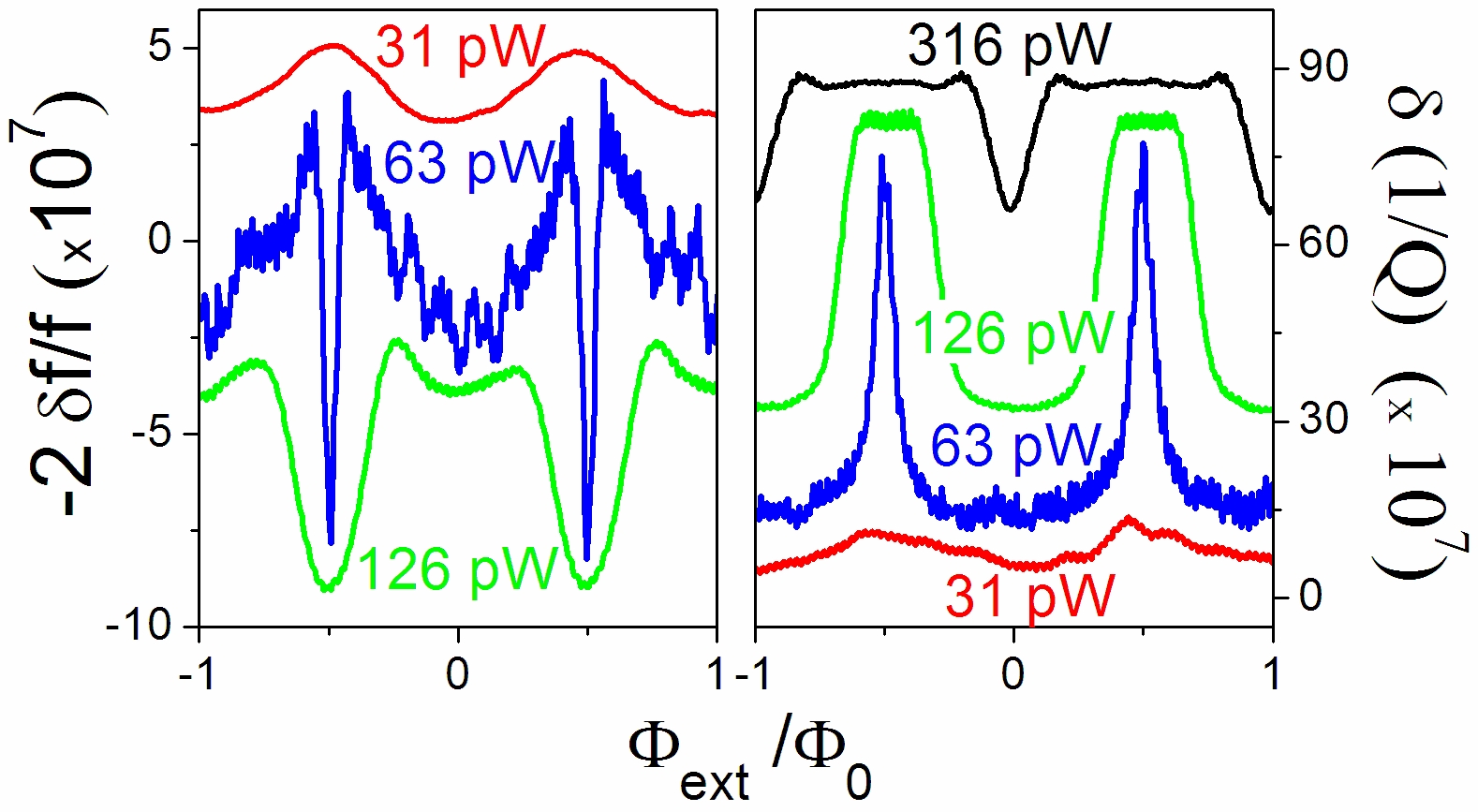}
    \caption{Non linear regime. Flux dependent dissipative and non-dissipative responses   for different excitation microwave powers at $T=0.7K$. This data was taken for the same N Au wire inbedded in a larger  ring than the previous measurements. }
    \label{Fig4}
\end{figure}
Moreover   the relaxation of the Andreev levels population   is expected to be  governed  by the inelastic electron scattering time \cite{blanter96} identical to the Nyquist dephasing time \cite{texier2006} in a finite size  quasi 1D wire. ( Low-energy electrons in the N wire    are totally Andreev-reflected from  the superconducting interfaces).  In the limit corresponding to our experimental situation   where $T \ll gE_{T}/k_B\simeq 1000K$, $\tau_{in} (T)= \tau_\varphi (T) = g \hbar/k_B T$  which is of the order of 100 ns i.e.   much longer that the time scales  (ns range) investigated in this experiment.   The characteristic time $\tau $ entering in the fits in Fig. 3   must therefore describe a different and faster  dynamics than the   equilibration of the population of Andreev levels described above.  The breakdown of the adiabatic approximation when Andreev levels do not follow the phase oscillations is expected to occur when  $\omega$  is larger than the inverse  Ginzburg Landau  relaxation time estimated to be $\tau_{GL}= (\pi^2/4)\tau_D $  in a wire of finite length \cite{Warlaumont79, skocpol}. This   justifies  to fit  the experimental results  at $\omega \gg 1/\tau_{in}$ with, 
  \begin {equation}
   \chi(\omega,\varphi,T)= \chi_d(\varphi,T) /(1+i\omega \tau).
 \label{eqdebye2}
 \end{equation}
   and $\tau$ expected to be of the order of $\tau_D$. We show in Fig. 2 that the experimental flux dependence of  $\delta\chi'$ and $\delta \chi''$ 
 at 0.55, 0.67 and  1K  and 2 different frequencies are well reproduced by Eq.(\ref{eqdebye2})   with $\tau=0.6ns$ as determined previously and  $\chi_d(\varphi)$  calculated  from expression (\ref {eqchid}). We have used a  unique rescaling factor   of the order of 0.7  for all the data  which can be attributed to  errors in the estimation of the mutual coupling between the ring and the resonator. 
     Recent results  \cite{Tikhonov} on the finite frequency  response of a SINIS junction where the normal mesoscopic wire is  isolated from the electrodes by an insulating barrier lead to  a frequency dependent response with similar flux dependences for $\chi'$  and  $\chi''$, as observed here and a characteristic time  given by $\tau_D$.  The extension of this work to our experimental configuration with highly transmitting N/S interfaces is in progress \cite{virtanen}. Finally  pair breaking induced by residual  magnetic impurities could also be important and  needs to be considered. 
 
 In conclusion we have found that the high frequency current response at GHz frequencies of NS rings is dramatically different from the flux derivative of the Josephson current. The harmonics content of the dc flux dependence is well understood as a result of the freezing of   the populations of the Andreev states which cannot follow the time dependent flux. The physical origin of the nanosecond time scale responsible for the observed dissipative response and the  associated high frequency decrease of the in phase response seems to be related to the diffusion time through the normal part of the ring. 

We acknowledge  help of A.Kasumov and F.Fortuna for fabrication using the FIB as well as useful discussions with R. Deblock, M. Aprili, B. Reulet and L. Glazman.


\begin{thebibliography}{99}
\bibitem{icdeT} We have checked that the condition $\beta=1$ is consistent with the calculated geometrical  ring inductance, and the critical current at this temperature.  We have determined $I_c(T)$ from  the  temperature dependence of amplitude of the hysteresis on the flux axis. The Thouless energy was  then determined from the expression of $I_c(T)$ calculated from  P. Dubos et al., Phys. Rev. {\bf B 63}, 064502 (2001), with the ratio $\Delta/ E_{Th} =84$ where $\Delta$ is the gap of the superconducting electrodes.
\end{thebibliography}
\end{document}